\documentclass[useAMS,usenatbib]{mn2e}
\usepackage{color}
\usepackage{graphicx}
\usepackage{epstopdf}
\usepackage{amsmath}

\usepackage{times}
\title[]{Structure of a hot accretion flow in the presence of outflow and convection with large ordered magnetic field}
\author[Maryam Ghasemnezhad]{Maryam Ghasemnezhad  \thanks{E-mail:m.ghasemnezhad@uk.ac.ir; m$_{-}$ghasemnezhad2005@yahoo.com}\\
 Faculty of physics, Shahid Bahonar University of Kerman, Kerman, Iran}

\date{}
\begin{document}
\pagerange{\pageref{firstpage}--\pageref{lastpage}} \pubyear{2017}
\maketitle \label{firstpage}
\begin{abstract}
Hydrodynamics and magnetohydrodynamic simulations of hot accretion flow have indicated that there is an inward decreasing of mass accretion rate with decreasing radius. Consequently, we have a flatter density profile ($\rho \propto r^{-p}$ with $p\propto 1.5-s$ in the case of $\dot{M}\propto r^{s}$, $0 \leq s<1$) compared to the case of a constant accretion rate (($\rho \propto r^{-\frac{3}{2}}$). In order to describe this behavior two models have been proposed: inflow-outflow solution and convection-dominated accretion flows. We have studied the properties of a hot accretion flow in the presence of convection, large scale magnetic field and outflow. We consider an axisymmetric viscous flow in the steady state. We used the self-similar solutions to solve the 1.5 dimensional inflow-outflow equations. We have assumed that the convection as a free parameter in our model for simplicity. We have considered two components of magnetic field (toroidal and vertical) in this paper. We have shown that the strong convection makes the inflow accretes and rotates slower while it becomes hotter and thicker. We have found
that the thickness of the disc deviates from non-convective solutions obviously. We have represented that two components of magnetic field have the opposite effects on the thickness of the disc and similar effects on the radial and angular velocities of the flow.
\end{abstract}

\section{INTRODUCTION}
The optically thin advection dominated accretion flows (ADAFs) belong to the hot regime of accretion flow. These flows proposed by Ichimaru (1977) and Rees et al. (1982). Then this type of solutions have been developed by Narayan \& Yi (1994, 1955) and Abramowicz et al. (1995) where successfully describe the general features of hot accretion flows in the low-luminosity accreting black holes and the hard X-ray emission seen in some black hole binary sources (Yuan \& Narayan 2014 for review). In contrast to the standard model, the heat generated by viscosity will be advected inwards with the flow instead of being radiated. These flows are radiatively inefficient because of the long cooling time.

Recent observations have revealed strong evidence for the presence wind or uncollimated outflow in our galactic center, Sg$A^{\star}$ (Wang et al. 2013). In hot accretion flows, the presence of wind helps to explain many observation features of hot accretion flows, such as Fermi bubbles in the Galactic Centre (Mou et al. 2014 ). Also it is important to study the nature of winds in hot accretion flows because it can strongly affect the structure of the inflowing gas, and can be a powerful contributor to AGN feedback.
\\
On the other hand, hydrodynamical (HD) and magneto-hydrodynamical (MHD) simulations have confirmed the existence of outflow in hot ADAF models (Li \& Ostriker et al. 2013, Sadowski et al. 2016, Bu et al. 2016a, 2016b, Ohsuga et al. 2005, 2009, Narayan et al. 2012, Yuan et al. 2012b, 2015). These simulations have found that the mass accretion rate is not constant and decreases inward in comparison to the originally ADAF solutions. In order to explain this inconsistency two alternative models have been proposed. The first one is the adiabatic inflow-outflow solution (ADIOS; Blandford \& Begelman 1999, 2004; Begelman 2012;  Jiao \& Wu 2011 for related works.) and the second one is the convection-dominated accretion flow (CDAF; Narayan et al. 2000; Quataert \& Gruzinov 2000).  In ADIOS model, the inward decrease of inflow rate is due to the loss of gas in the form of outflow/wind but in CDAFs, the inward decrease of mass accretion rate is because of the convection. 
 The difference between two scenarios of ADAFs would be seen by numerical simulations. In ADIOS model, Blandford \& Begelman (1999) proposed that the radial dependency of the mass accretion rate is as $\dot{M}\propto r^{s}$ with  $0 \leq s<1$ ( which the index $s$ is a constant and show the strong of the wind parameter ).\\ In the CDAFs model, the accretion gas is convectively unstable and is locked in convective eddies. The convective bubbles originate in the innermost of ADAFs and transport energy outward. Also the convective bubbles transport angular momentum inward and suppress the accretion process. The motion of the flow is assumed to be dominated by convective turbulence.
There is some possible mechanisms for the origin of outflow in hot ADAF models such as the positive Bernoulli parameter, convection and large scale ordered magnetic fields ( e.g., the magnetic centrifugal force)(Narayan \& Yi 1994, 1955, Blandford \& Begelman 1999, Yuan et al. 2015, Yuan \& Narayan 2014). \\

Yuan et al. (2012b) have found that the origin of outflow in HD accretion flow is the buoyancy force which arises because of the convective instability and in the MHD accretion flow is the combination of centrifugal gas and magnetic pressure gradient forces which arise because of the magnetic field.  Yuan et al. (2012b) have shown that the MHD accretion flows are convectively stable in the most of the regions of the flow and this implies that for MHD flows, the CDAF model at least can not be applied. So the magnetorotational instability (MRI) controls the dynamics of MHD flows. Also Narayan et al. (2002) performed a linear MHD stability analysis of MHD flows and found that in general both MRI and convection modes exist. They have argued that when the flow is convectively unstable, the result depends on the wavelength of the perturbation. The long-wavelength modes of the instability are found to be independent of the magnetic field and are convective modes. So the CDAF models can be applied for MHD flows. \\Although the results of Yuan et al. (2012b) seem to favor the ADIOS models, it is possible that the fact involves the combination of the two models of ADAFs (Yuan \& Narayan 2014). There are some debates about whether convection exists in MHD accretion flow (e.g. Stone et al. 1999, Narayan et al. 2000). Of course, some uncertainties still exist and we can't conclude the non-existence of convection (see discussion in Yuan \& Narayan 2014). Thus it is still feasible to study convection and it can be valuable work to study outflow, magnetic field and convection in hot accretion flow.
 \\
 The importance and presence of magnetic fields in the accretion discs are generally accepted. The magnetic field has several effects such as the formation of the wind/jet (Yuan et al. 2015), suppressing the convective instabilities of hot accretion flow (Yuan et al. 2012b and Narayan et al. 2012) and the transferring the angular momentum through the MRI process in accretion flow (Balbus \& Hawley 1998). In order to study hot accretion flow analytically several attempts have been done including one-dimensional, 1.5 dimensional and two-dimensional assumptions (Kaburaki 2000, Akizuki \& Fukue 2006, Abbassi et al. 2010, Ghasemnezhad et al. 2012, 2013, Ghasemnezhad \& Abbassi 2016, Samadi \& Abbassi 2016, Ghasemnezhad \& Abbassi 2016, Samadi et al. 2014, Bu et al. 2009, Zhang \& Dai 2008). \\
 The numerical MHD simulations have shown that a large scale magnetic field can exist in the inner regions of hot accretion flows (Hirose et al. 2004).  Figure 6. in Hirose et al. (2004) has shown that the toroidal component of magnetic field ($B_{\phi}$) is dominant in the inner regions of the main body of accretion flow but near the pole, the poloidal component of magnetic field in the vertical direction is dominant ($B_{z}$). \\
 Bu et al. (2009) studied self similar solutions of hot ADAFs in the presence of ordered large scale magnetic
field and outflow in the absence of the convection effect. Their solutions showed the dynamical structure of ADAFs modified
significantly when ordered magnetic field and outflow existed. In this paper, we
develop Bu et al. (2009) and Xi \& Yuan (2008) solutions by considering the
effect of convection in the 1.5 dimensional inflow-outflow equations. In this work, similar to the method of Bu et al. (2009) and Xi \& Yuan (2008), we have investigated the dynamical effect of outflow by considering the interchange of mass, angular momentum and energy between inflow and outflow. \\
Hypothesis of the model and relevant equations are developed in
Section 2. Self-similar solutions are presented in Section 3. We
show the result in Section 4.

\section{Basic Equations}
The main goal of present paper is studying the possible effect of convection on the structure of hot accretion flow in the presence of large scale magnetic field. Interchange of mass, momentum (radial and angular) and energy between inflow and outflow have been considered. Following  Bu et al. (2009) and Xi \& Yuan (2008) we formulate the equations of our system using the self-similar method. We consider a steady state ($\frac{\partial}{\partial t}=0$) and axisymmetric ($\frac{\partial}{\partial \varphi}=0$) hot accretion flow in cylindrical coordinates ($r,\varphi,z$). We have integrated all MHD equations in the vertical direction so we write the 1.5 dimensional equations instead of full two dimensional equations. For simplicity, we ignore the self-gravity of the discs and the general relativistic effects. The Newtonian gravity is used in the radial direction. The convection, outflow and magnetic field transfer the energy and angular momentum in the disc. \\
 Following Bu et al. (2009), lovelace et al. (1994) and Hirose et al. (2004) assumptions for studying the large scale magnetic field, even symmetry with respect to the equatorial plane for $B_{z}$ and odd symmetry for $B_{\varphi}$ have been adopted. Also the vertical gradient of $B_{z}$ is neglected.\\
 The magnetic field structure has been studied numerically in the different parts of the flow (main disc body , the inner torus and plunging region, the corona envelope and the funnel part) by Hirose et al. (2004). The numerical MHD simulations done by Hirose et al. (2004) have shown that the magnetic field includes two components as: the large scale component and the small scale component (turbulent component). According to these simulations, the turbulence and the differential rotation effects control the magnetic field geometry in the main body of the disc. The former bends the field line in all directions and the latter removes radial field line into toroidal field line. So the magnetic field configuration of the main disc body is tangled toroidal field and turbulent field. In the plunging and inner regions, the flow changes from turbulence dominated to spiraling inflow. Also in the corona, the magnetic field is purely toroidal without tangles. Although the large scale poloidal magnetic field only exists in the funnel part. Therefore the toroidal magnetic field governs in the main body of the flow and in its inner parts while the funnel part is governed by a poloidal magnetic fields (which is mostly in the vertical direction $z$). Thus we consider two components of large scale magnetic field $B_{z}$ and $B_{\varphi}$ and we suppose that $B_{r}=0$. We used the usual $\alpha$ prescription for the turbulent viscosity that consists all the effects of small scale (turbulence) component of magnetic field. In this paper we consider the even $z-$ symmetry field ($ B_{\varphi}(r,z)=- B_{\varphi}(r,-z) $, $B_{z}(r,z)=+ B_{z}(r,-z)$) and $B_{z}$ is in the form of even function of $z$. We can write the magnetic field as follows:
\begin{equation} 
\vec{B}=\vec{B_{p}}(r,z)+\vec{B_{\varphi}}(r,z)\hat{\varphi}
\end{equation}
where $B_{p}$ and $B_{\varphi}$ are the poloidal and toroidal magnetic field components. We can express that $\Delta B_{Z} = \frac{H}{r} (B_{r})_{H}$ from the magnetic divergence-free condition and by assuming two conditions as $\frac{H}{r} \leq 1$ and $(B_{r})_{H} \leq B_{z}$, it follows that $\Delta B_{Z} << B_{Z}$, that is, we have neglected the variations of $B_{z}$ with $z$ in the disc.
  So we have:
\begin{equation}
B_{r}=0
\end{equation}
\begin{equation}
B_{\varphi}=B_{0}\frac{z}{H}, \rightarrow B_{\varphi,z=H}= -B_{\varphi,z=-H}
\end{equation}

where $H$ is the half-thickness of the accretion disc and $B_{0}$ to be $B_{0}=B_{\varphi,z=H}$. Xie \& Yuan (2008) have considered the accretion flows consist of two parts: inflow and outflow. They have supposed that the inflow is the vertically hydrostatic equilibrium $v_{z}=0$ while outflows are launched from the surface of inflow ($z=H$) with their own velocity as $v_{z,w}, v_{r,w}, v_{\varphi,w}$. Therefore in the surface of disc there is the discontinuity of mass, momentum and energy between inflow and outflow.
   
 We can write the equations of conservation of mass, momentum which are as follows (Bu et al. 2009):
 \begin{equation}
	\frac{d\dot{M}(r)}{dr}=\eta_{1}4\pi r \rho v_{z,w}.
\end{equation}
where the mass accretion rate is defined as $\dot{M}(r)=-2 \pi r v_{r}\rho $. Also $\rho$ and $\eta_{1}$ are the density of flow and the density ratio of outflow and inflow ( $\eta_{1}=\frac{\rho_{w}}{\rho}\cong 0.71$) respectively (Xie \& Yuan 2008).

\begin{displaymath}
v_{r}\frac{d v_{r}}{dr}+\frac{1}{2\pi r \Sigma}\frac{d \dot{M}(r)}{dr}(v_{r,w}-v_{r})=\frac{v^{2}_{\varphi}}{r}-\frac{GM}{r^{2}}
\end{displaymath}
\begin{equation}
-\frac{1}{\Sigma}\frac{d(\Sigma c^{2}_{s})}{dr}-\frac{1}{4\pi \Sigma}[\frac{d}{dr}(H B^{2}_{z})
+\frac{1}{3}\frac{d}{dr}(H B^{2}_{0})+\frac{2}{3}\frac{B^{2}_{0}}{r}H],
\end{equation}
\begin{displaymath}
\frac{\Sigma v_{r}}{r}\frac{ d rv_{\varphi}}{dr}+\frac{1}{2\pi r}\frac{d\dot{M}(r)}{dr}(v_{\varphi,w}-v_{\varphi})= \frac{1}{r^{2}}\frac{d J_{vis}}{dr}
 \end{displaymath}
 \begin{equation}
\frac{1}{r^{2}}\frac{d J_{con}}{dr}-\frac{1}{2\pi}B_{Z}B_{0},
 \end{equation}
 \begin{equation}
 \frac{GM}{r^{3}}=(1+\beta_{1})c^{2}_{s}
 \end{equation}
 where $\Sigma \equiv \int \rho dz$ is the vertically integrated density. All variables have their own usual meanings. $v_{r}$, $v_{\varphi}$, $\Omega$, $G$, $J_{vis}=r^{3}\nu \Sigma \frac{d\Omega}{dr}$, $J_{con}=r \nu_{con} \Sigma \frac{d(r^{2}\Omega)}{dr}$ are the radial speed, the rotational speed, the angular speed of inflow, the gravitational constant, the viscous and convective angular momentum
fluxes respectively. The convective angular momentum flux is oriented down the specific
angular momentum gradient. Despite the theoretical studies (Ryu \& Goodman 1992, Kumar et al. 1995) and numerical simulations (Stone \& Balbus 1996) have shown that the convection transfers angular momentum inward, Narayan et al. (2000) have studied the transport of angular momentum by convection. They have mentioned that is a complex subject and there is no general agreement on how it works. There are two possibilities for transportation of angular momentum by convection (Narayan et al. 2000). One possibility is the convection behaves like usual viscosity and the convective angular momentum flux is oriented down the angular velocity gradient $(J_{con}=r^{3}\nu_{con}\Sigma\frac{d\Omega}{dr})$ and the convection tries to drive a system towards a state of uniform rotation like viscosity. Another possibility is that the convective angular momentum flux is oriented down the specific angular momentum gradient. So that convection tries to drive a system towards a state of uniform specific angular momentum and so the convection carries angular momentum inward and suppresses accretion process $(J_{con}=r \nu_{con}\Sigma\frac{d}{dr}(r^{2}\Omega))$. Narayan et al. (2000) have shown that the general form of the convective angular momentum flux is as ($J_{con}=\nu_{con}\Sigma r^{3\frac{(1+g)}{2}}\frac{d}{dr}(\Omega r^{3\frac{(1-g)}{2}})$). The convection transports angular momentum inward (or outward) depending on $g$ parameter, for $g < 0$ (or $g> 0$), and the specific case $g = 0$ corresponds to zero angular momentum transport (Narayan et al. 2000). In this paper we consider the convective angular momentum flux as ($ J_{con}=r\nu_{con}\Sigma\frac{d}{dr}(r^{2}\Omega)$) where it is related to $g=-\frac{1}{3}$ and shows the inward transport of angular momentum. The parameters $\nu$ and $\nu_{con}$ are the kinematic viscosity coefficient and convective diffusion coefficient. We have formalized the convective diffusion similar to viscosity turbulence by Shakura \& Sunyaev (1973). So we have:
\begin{equation}
 \nu=\alpha c_{s}H
 \end{equation}
\begin{equation}
 \nu_{con}=\alpha_{c} c_{s}H
 \end{equation}
 where $\alpha$, $\alpha_{c}$ are the dimensionless viscous and convective parameters. We consider the convective coefficient $\alpha_{c}$ as a
free parameter to discuss the effects of convection for simplicity. \\
Also $\beta_{1}=\frac{(B_{\varphi,z=H})^{2}/8\pi}{\rho c^{2}_{s}}$ and $\beta_{2}=\frac{(B_{z,z=H})^{2}/8\pi}{\rho c^{2}_{s}}$ are the ratio of magnetic pressure (azimuthal, vertical) to gas pressure.
For hot accretion flow, the $\beta_{1,2}$ can change in the range $0.01-1$ (De Villiers et al. 2003).\\
 The induction equation of the magnetic field is:
\begin{equation}
	\frac{\partial \vec{B}}{\partial t}=\vec{\nabla}\times (\vec{v}\times \vec{B}-\frac{4\pi}{c}\eta\vec{J}).
\end{equation}
 where $\vec{J}=\frac{c}{4\pi}\vec{\nabla}\times \vec{B}$ is the current density. The induction equation is the field escaping/creating rate due to magnetic instability or dynamo effect. For the steady state accretion flow, we neglect the dynamo effect. In the energy equation we assumed a balance between the heating due to viscosity and cooling due to advection, convection and radiation as:
\begin{equation}
 Q^{-}_{adv}+Q^{-}_{rad}+Q^{-}_{con}=Q^{+}_{diss} 
 \end{equation}
where $Q^{-}_{adv}=\rho (\frac{d e}{dt}-\frac{p}{\rho^{2}}\frac{d\rho}{dt})$ is advection cooling in ADAFs. $e$ is the gas internal energy $(e=\frac{c^{2}_{s}}{\gamma-1})$ and $\gamma$ is the specific heat ratio. Also we consider $Q^{+}_{diss}-Q^{-}_{rad}=fQ^{+}_{diss}$ in ADAFs (where $f$, the advection parameter, is defined  by Narayan \& Yi (1994)). Also the viscous heating is defined as $Q_{diss}= f(\nu+\frac{1}{3}\nu_{con})\Sigma r^{2} (\frac{d \Omega}{dr})^{2} $. The outward energy flow by convection is $ Q^{-}_{con}=- \vec{\nabla}\cdot \vec{F_{c}}$ where $F_{con}=-\nu_{con} \frac{1}{\gamma-1}\frac{d c^{2}_{s}}{dr}-\frac{c^{2}_{s}}{\rho}\frac{d \rho}{dr}$. So the energy equation is written by considering the above definitions and the specific internal energy of inflow $\epsilon$ and outflow $\epsilon_{w}$ in the surface of the disc which is as follows:

\begin{displaymath}
 \frac{\Sigma v_{r}}{\gamma-1}\frac{d c^{2}_{s}}{dr}-2Hc^{2}_{s}v_{r}\frac{d\rho}{dr}+\frac{1}{2\pi r}\frac{d\dot{M}(r)}{dr}(\epsilon_{w}-\epsilon)-
\end{displaymath} 
\begin{displaymath}
\frac{1}{r}\frac{d}{dr}(r\nu_{con}\frac{\Sigma}{\gamma-1}\frac{d c^{2}_{s}}{dr}-2Hr\nu_{con}c^{2}_{s}\frac{d\rho}{dr})-
\end{displaymath}
 \begin{equation}
 \frac{\nu \rho}{\gamma-1}\frac{d c^{2}_{s}}{dr}+\nu_{con}c^{2}_{s}\frac{d\rho}{dr} = f\Sigma (\alpha-\frac{1}{3}\alpha_{c}) c_{s}H r^{2}(\frac{d \Omega}{dr})^{2}
 \end{equation}

  \section{Self-Similar Solutions}
  For solving the above equations, we present the self similar method.  The self similar solutions are a dimensional analysis and powerful techniques to give an approximate solution for differential MHD equations. This method has been adopted several times to solve ADAFs governing equations (Narayan \& Yi (1994) and Bu et al. (2009)). Following Bu et al. (2009) we have:
  \begin{equation}
\Sigma=\Sigma_{0} (\frac{r}{r_{out}})^{s}
\end{equation}
\begin{equation}
v_r(r)=-c_1 \alpha \sqrt{\frac{G M}{r_{out}}}(\frac{r}{r_{out}})^{-\frac{1}{2}}
\end{equation}
\begin{equation}
V_\varphi(r)= r \Omega (r) =c_2 \sqrt{\frac{G M}{r_{out}}}(\frac{r}{r_{out}})^{-\frac{1}{2}}
\end{equation}
\begin{equation}
c_\mathrm{s}^{2}=c_3 \frac{G M}{r_{out}}(\frac{r}{r_{out}})^{-1}
\end{equation}
  The accretion rate decreases inward as:
  \begin{equation}
  \dot{M}(r)=\dot{M}(r_{out})(\frac{r}{r_{out}})^{s+\frac{1}{2}}
  \end{equation}
  \begin{equation}
  \dot{M}(r)=2\pi c_{1}\alpha \Sigma_{0}\sqrt{GM} r^{\frac{1}{2}}_{out}(\frac{r}{r_{out}})^{s+\frac{1}{2}}
  \end{equation}
  where $\dot{M}(r_{out})$ and $\Sigma_{0}$ are the mass inflow rate and the surface density at the outer boundary ($r_{out}$). The index $s$ is constant and for $s=-\frac{1}{2}$ we have solutions without outflow/wind.\\
  By substituting the above self-similar solutions in to the dynamical equations of the
system, we obtain the following system of dimensionless equations, to be solved for $c_{1}$, $c_{2}$ and $c_{3}$:
\begin{displaymath}
-\frac{1}{2}c^{2}_{1}\alpha^{2}-(s+\frac{1}{2})(\zeta_{1}-1)c^{2}_{1}\alpha^{2}=c^{2}_{2}-1-(s-1)c_{3}
\end{displaymath}
\begin{equation}
-\beta_{2}(s-1)c_{3}-\frac{1}{3}\beta_{1}(s-1)c_{3}-\frac{2}{3}\beta_{1}c_{3}
\end{equation}
\begin{displaymath}
\alpha c_{1} c_{2}-2\alpha (s+\frac{1}{2})(\zeta_{2}-1)c_{1}c_{2}=3\alpha c_{3}c_{2}(s+1)\sqrt{1+\beta_{1}}
\end{displaymath}
\begin{equation}
-\alpha_{c}c_{3}c_{2}\sqrt{1+\beta_{1}}(s+1)+4\sqrt{\frac{c_{3}\beta_{1}\beta_{2}}{1+\beta_{1}}}
\end{equation}
\begin{equation}
\frac{H}{r}=\sqrt{(1+\beta_{1})c_{3}}
\end{equation}

\begin{displaymath}
   c_{1}\alpha[\frac{1}{\gamma-1}+(s-1)+\frac{(s+\frac{1}{2})(\zeta_{3}-1)}{\gamma-1}]+
\end{displaymath}
\begin{displaymath}
\alpha_{c}\sqrt{1+\beta_{1}}c_{3}(s+\frac{1}{2})[\frac{1}{\gamma-1}+(s-1)]+\frac{\alpha_{c}\sqrt{c_{3}}}{2}[\frac{1}{\gamma-1}+(s-1)]=
\end{displaymath}
\begin{equation}
\frac{9}{4} f c^{2}_{2}\sqrt{1+\beta_{1}}(\alpha+\frac{1}{3}\alpha_{c})
\end{equation}

 By introducing  $\zeta_{1,2}$ and $\zeta_{3}$ we are able to evaluate the radial and azimuthal velocities and internal energy of the outflow in terms of inflow: $v_{r,w}=\zeta_{1}v_{r},v_{\varphi,w}=\zeta_{2}v_{\varphi}$ and $\epsilon_{w}=\zeta_{3} \epsilon$. When $\zeta_{3}>1$, the outflow is extra cooling rate for the inflow and is the extra heating rate for the inflow when $\zeta_{3}<1$.
 
The second terms on the left hand side of the Eq.(20) is angular momentum transfer between inflow and outflow (the angular momentum is taken away from the inflow by the outflow when $\zeta_{2}>1$ or deposited into the inflow by the outflow if $\zeta_{2}<1$)(Xie \& Yuan 2008 and Bu et al. 2009). Also on the right hand side of this equation we have the angular momentum transfer outward by turbulent viscosity and the large scale magnetic field. The convection diffusion in this equation transfers angular momentum inward similar to outflow in the case of ${\zeta_{2}<1}$. The ratio of the inward transfer angular momentum by convection and outflow is calculated by defining the new parameter $c_{4}$ as:
\begin{equation}
c_{4}=\frac{\alpha_{c}c_{3}c_{2}\sqrt{1+\beta_{1}}(s+1)}{-2\alpha (s+\frac{1}{2})(\zeta_{2}-1)c_{1}c_{2}}
\end{equation}
\\
  By solving the Eq.(19-22) numerically for given values of $\alpha$,$\alpha_{c}$, $f$, $s$, $\zeta_{1,2,3}$ and $\beta_{1,2}$, we will study the structure of the hot flow in the presence of convection diffusion, the large ordered magnetic field and outflow. Our results reduce to the solutions of
Bu et al. (2009) without the convection diffusion ($\alpha_{c}=0$). 
  \section{Results}
  We are interested to consider the effects of convection as a free parameter, outflow and large ordered magnetic field on the structure of accretion disc. The constant values of $\alpha=0.1$, $s=0.5$ (that means $\dot{M}(r)\propto r$, which are related to a strong wind), $f=1$ and $\gamma=\frac{4}{3}$ are set in all figures. In three panels of Fig. 1, we investigate the effect of convection parameter $\alpha_{c}$ on the dynamics of hot accretion flows. Fig. 1 shows the self-similar coefficient $c_{1}$
(the radial velocity), $c_{2}$ (the rotational velocity) and $\frac{H}{r}$ (the relative thickness) as functions of the toroidal component of magnetic field $B_{\varphi}(\beta_{1})$ for different values of the convection parameter $\alpha_{c}$
(= 0.01, 0.06, 0.1). We have compared these solutions with Bu et al. (2009) which their solutions are related to $\alpha_{c}=0$. By increasing the convection parameter $\alpha_{c}$ from $0.01$ to $0.1$, we can see that the radial and angular velocities decrease while the thickness of the disc (or the sound speed) increases. It can be easily understood since the convection carries angular momentum inward while the turbulence viscosity transport angular momentum outward and so viscosity makes the mass flow inward. Therefore the convection will make the flow accretes slowly and suppress the accretion process. Generally convection tries to drive the disk towards a state of uniform specific angular momentum. So the radial and toroidal velocities of the flow decrease by increasing of convection effect. Also the convective diffusion, is a dissipative process like viscosity which makes the flow becomes hotter and thicker. Furthermore, the stronger the toroidal magnetic field is, the faster the accretion flow falls and rotates and the more thickness will be in case $\beta_{1}> 0.2$. This is because, the magnetic field can transport angular momentum outward. A magnetized
disk must rotate faster and accrete more inward than when there
is nonmagnetic field present because of the effect of magnetic stress force ($-\frac{1}{2}\beta_{1}c_{3}$). As can be seen, the sound speed decreases in the presence of large scale magnetic field. A magnetized disc must rotate faster and accrete more inward because of the effect of magnetic stress force (turbulence) (or a centripetal force). As a result, the centrifugal force (both the gas pressure gradient force ($\propto c_{3}$) and the magnetic pressure gradient force ($\propto c_{3}$)) decreases in the presence of large scale magnetic field since the temperature $(c_{3})$ will decrease for the wide range of $\beta_{1,2}$. So the disc becomes thicker by increasing the the toroidal component of magnetic field ($\beta_{1}$) (see Eq.(21)).
 These results are qualitatively consistent with results presented by Bu et al. (2009). Also we have found that only the thickness of the disc 
deviations from non-convective solutions obviously (Bu et al. (2009)). 
  We can see in Figs.1,2,3,4 the radial and rotational velocities decrease by adding the convection diffusion but the thickness of the disc increases. The effects of convection, vertical component of magnetic field and outflow parameters on the $c_{4}$ have been plotted in Fig. 2,3,4. As we
can see in Eq. (23), if $c_{4}< 1$, the dominant mechanism in inward angular momentum transport will be the outflow. We can see in Fig.2,3,4, $c_{4}<1$ and increases by increasing convection parameter (see Eq. (23), it can be understood obviously).  
 We have studied the effect of vertical component of magnetic field $B_{z}(\beta_{2})$ in the structure of hot flow. We have shown that by increasing the $\beta_{2}$, the thickness of the flow (the sound speed) decreases and then the centrifugal force $-\beta_{2}(s-1)c_{3}-\frac{1}{3}\beta_{1}(s-1)c_{3}$ decreases. As a result, the infalling velocity and the angular velocity will increase which is in a great agreement with Bu et al. (2009). As can be seen in Fig. 2, the convection contributor in inward angular momentum transfer decreases by adding the vertical component of magnetic field while in the lower value of $\beta_{2}=0.01$
  and in the high value of convection parameter $\alpha_{c}> 0.08$, the $c_{4}$ is bigger than unity. So the convection has a bigger role in the angular momentum transport inward with respect to the outflow.
  We have studied the effects of angular momentum and energy of outflow in the presence of convection in the parameter $c_{4}$ in two Fig. 3,4. We have shown that in Fig. 3, by increasing the angular momentum of outflow the hot flow accretes more, rotates faster and becomes thinner that is in the great agreement with Bu et al. (2009). This figure reveals
that increasing the angular momentum of outflow has obvious effect on
the $c_{4}$ parameter when the rotational velocity of wind is bigger than that of inflow. So the convection contributor in inward angular momentum decreased by adding the $\zeta_{2}=0.9, 1.1, 1.2$. 
  Finally, We have evaluated the role of outflow's energy in velocities and thickness of the flow. We have found that the thickness and radial velocity of the flow decreases when the energy is transfered outward by outflow but inflow rotates faster. Also the convection contributor in inward angular momentum decreased by adding the $\zeta_{3}$.
 
  \begin{figure*}
\centering
\includegraphics[width=138mm]{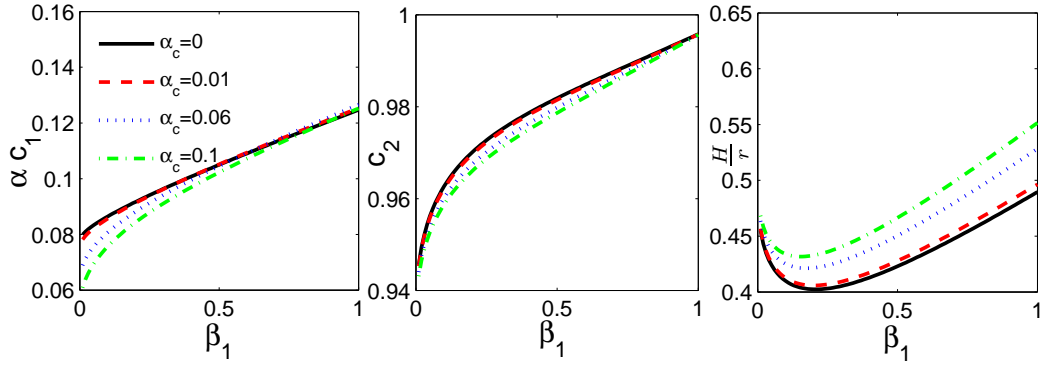}
\caption{ Numerical coefficient $\alpha c_1$,$c_2$ and $\frac{H}{r}$ as function of the toroidal component of magnetic field $\beta_{1}$ for several values of $\alpha_{c}$ (the convection parameter). For all panels we use $s=0.5$, $f=1$, $\gamma=1.33$, $\alpha=0.1$, $\zeta_{1,3}=1$, $\zeta_{2}=0.8$, $\beta_{2}=0.01$.}
\label{fig1}
\end{figure*}
  
 \begin{figure*}
\centering
\includegraphics[width=138mm]{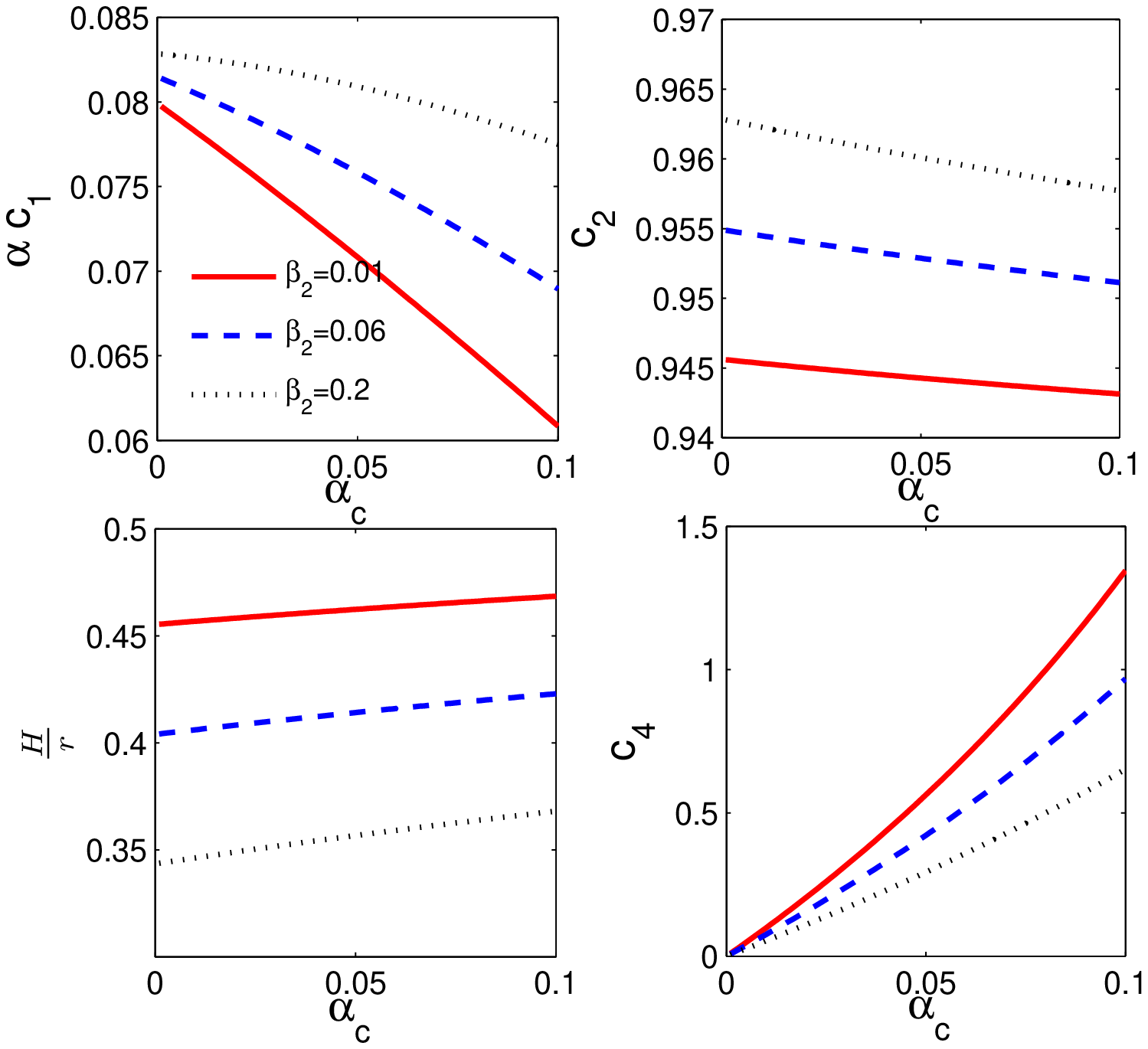}
\caption{ Numerical coefficient $\alpha c_1$,$c_2$, $\frac{H}{r}$ and $c_{4}$ as function of the convection parameter $\alpha_{c}$ for several values of $\beta_{2}$ (the vertical component of magnetic field ). For all panels we use $s=0.5$, $f=1$, $\gamma=1.33$, $\alpha=0.1$, $\zeta_{1,3}=1$, $\zeta_{2}=0.8$, $\beta_{1}=0.01$.}
\label{fig2}
\end{figure*}

\begin{figure*}
\centering
\includegraphics[width=138mm]{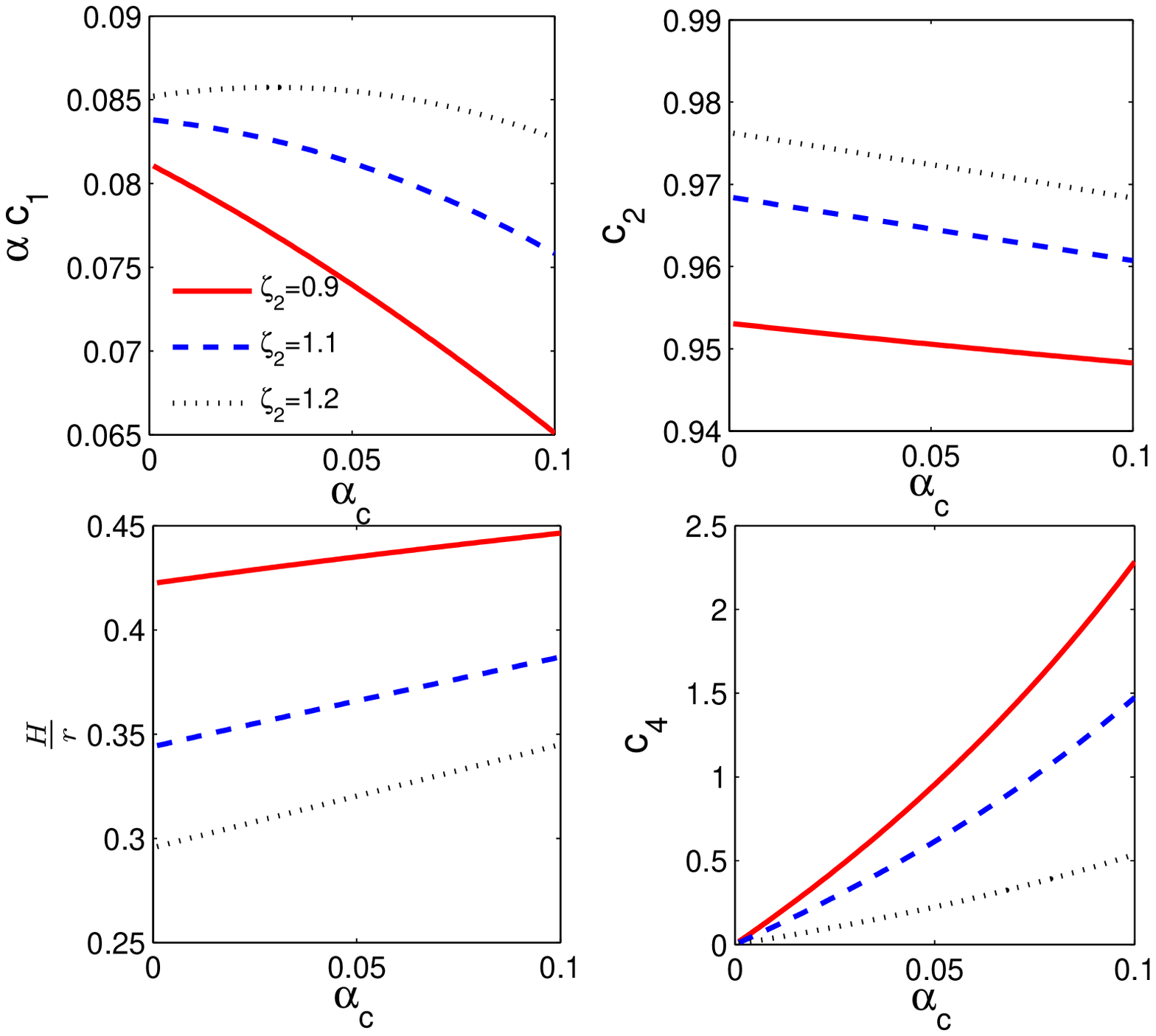} 
\caption{ Numerical coefficient $\alpha c_1$,$c_2$, $\frac{H}{r}$ and $c_{4}$ as function of the convection parameter $\alpha_{c}$ for several values of $\zeta_{2}$ . For all panels we use $s=0.5$, $f=1$, $\gamma=1.33$, $\alpha=0.1$, $\zeta_{1}=1$, $\zeta_{3}=1$, $\beta_{1,2}=0.01$.}
\label{fig3}
\end{figure*}

\begin{figure*}
\centering
\includegraphics[width=138mm]{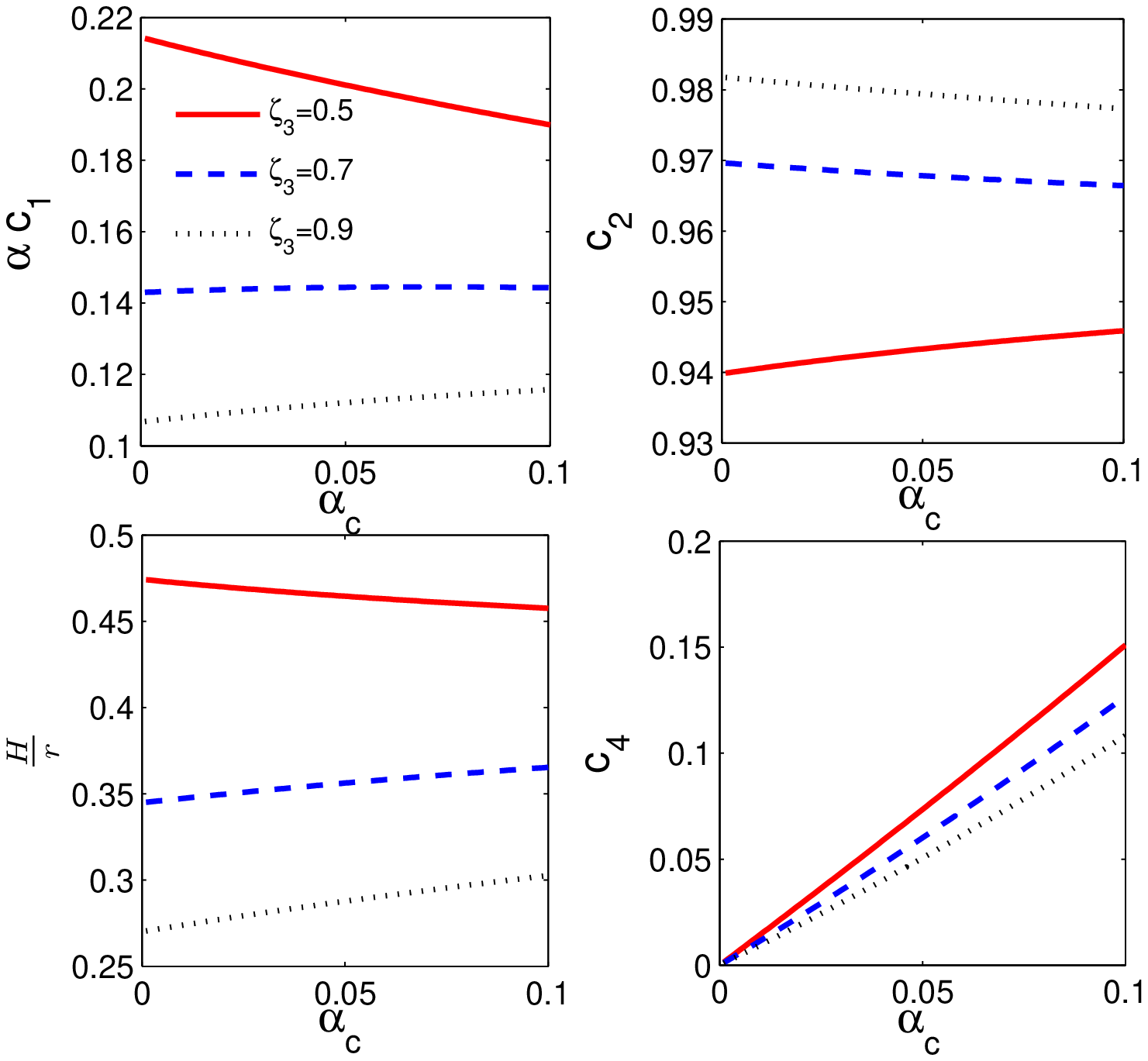}
\caption{ Numerical coefficient $\alpha c_1$,$c_2$, $\frac{H}{r}$ and $c_{4}$ as function of the convection parameter $\alpha_{c}$ for several values of $\zeta_{3}$. For all panels we use $s=0.5$, $f=1$, $\gamma=1.33$, $\alpha=0.1$, $\zeta_{1}=1$, $\zeta_{2}=0.5$ $\beta_{1,2}=0.2$.}
\label{fig4}
\end{figure*}
  \section{Summary and Conclusion}
  Numerical HD and MHD simulations have confirmed that there is a inward decreasing of mass accretion rate as $\dot{M}\propto r$. Both convection and outflow can be models for inward decreasing of mass accretion rate. So in this paper,
 we have studied the structure of hot accretion flow  in the presence of convection, large scale magnetic field and interchange of mass, momentum and energy between inflow and outflow by using the method of Bu et al. (2009) and Xi \& Yuan (2008).
  The large scale ordered magnetic field exists in the accretion disc and simulations have shown that the azimuthal and vertical components of magnetic field are important in the main body of the disc and near the pole respectively.\\  
 The self-similar method is used for solving the 1.5 dimensional, the steady state ($\frac{\partial}{\partial t}=0$) and axisymmetric ($\frac{\partial}{\partial \varphi}=0$) inflow-outflow equations to expand our understanding of the physics of the accretion discs around black hole. 
We have applied some limitations in our models for simplicity, for example we have ignored the self-gravity of the disc, the relativistic effects and we have used Newtonian gravity in the radial direction. We have compared our solutions in the presence of convection with non-convective solutions of Bu et al. (2009).\\
  We have found that the strong convection makes the inflow accretes and rotates slowly while it becomes hotter and thicker. We have found
that the thickness of the disc deviates from non-convective solutions obviously. We have represented that two components of magnetic field have the opposite effects on the thickness of the disc and similar effects on the radial and angular velocities of the flow.\\
 We have shown that the strong convection parameter causes the convection contribution in inward angular momentum transfer increases, while the strong vertical magnetic field, the angular momentum and energy of outflow decrease the convection contribution in inward angular momentum transfer.  The solutions showed that the specific energy and angular momentum of outflow and the vertical component of magnetic field have strong effects on the structure of hot accretion flow where these solutions are in a great agreement with Bu et al. (2009).
 Although that we have made some simplification and some limitations in order to solve equations numerically, our results show that convection can  affect structure of hot accretion flow which means in any realistic model this factor should be taken into account. 
 
 \section{ACKNOWLEDGEMENTS}
I am grateful to Shahram Abbassi for helpful comments on the paper. I also thank the referee for his/her thoughtful and constructive comments which greatly helped me to improve the paper.

\end{document}